\documentclass[prb,aps,twocolumn,showpacs]{revtex4}
\usepackage{bm}
\usepackage{epsfig}
\usepackage{amsmath}

\newcommand{\bpm}{\begin{pmatrix}}
\newcommand{\epm}{\end{pmatrix}}
\newcommand{\ba}{\begin{eqnarray}}
\newcommand{\ea}{\end{eqnarray}}

\begin{document}

\title{Effects of charge inhomogeneities on elementary excitations in La$_{2-x}$Sr$_{x}$CuO$_{4}$}

\author{S. R. Park$^1$, A. Hamann$^2$, L. Pintschovius$^2$, D. Lamago$^{2,3}$, G. Khaliullin$^{4}$, M. Fujita$^5$, K. Yamada$^6$, G. D. Gu$^7$, J. M. Tranquada$^7$, D. Reznik$^{1}$}

\email[Electronic address:$~~$]{dmitry.reznik@colorado.edu}

\affiliation{$^1$Department of physics, University of Colorado at
Boulder, Boulder, Colorado, USA}

\affiliation{$^2$Karlsruher Institute f\"ur Technologie, Institute f\"ur Festk\"orperphysik, P.O. Box 3640, D-76021 Karlsruhe, Germany}

\affiliation{$^3$Laboratoire Leon Brillouin, CEA-Saclay, F-91191 Gif sur Yvette Cedex, France}

\affiliation{$^4$Max-Planck-Institut f\"ur Festk\"orperforschung, Heisenbergstrasse 1, D-70569 Stuttgart, Germany}

\affiliation{$^5$Institute for Materials Research, Tohoku University, Senda, Miyagi 980-8577, Japan}

\affiliation{$^6$IWPI Research Center, Advanced Institute for Materials Research, Tohoku University, Sendai 980-8577, Japan }

\affiliation{$^7$Condensed Matter Physics $\&$ Materials Science Department, Brookhaven National Laboratory, Upton, New York 11973, USA}

\date{\today}

\begin{abstract}
Purely local experimental probes of many copper oxide superconductors show that their electronic states are inhomogeneous in real space. For example, scanning tunneling spectroscopic (STS) imaging shows strong variations in real space, and according to nuclear quadrupole resonance (NQR) studies the charge distribution in the bulk varies on the nanoscale.  However, the analysis of the experimental results utilizing spatially-averaged probes often ignores this fact. We have performed a detailed investigation of the doping-dependence of the energy and line width and position of the zone-boundary Cu-O bond-stretching vibration in La$_{2-x}$Sr$_{x}$CuO$_{4}$ by inelastic neutron scattering. Both our new results as well as previously reported angle-dependent momentum widths of the electronic spectral function detected by angle-resolved photoemission can be reproduced by including the same distribution of local environments extracted from the NQR analysis.
\end{abstract}
\pacs{63.20.-e, 74.25.Kc, 63.50.-x} \maketitle

\section{Introduction}
It is by now recognized that charge inhomogeneity is an important
aspect of copper oxide superconductors. Undoped copper oxygen
planes are Mott insulators due to a strong on-site Coulomb
interaction. They become metallic and exhibit high-temperature superconductivity (HTSC), when they are
doped.\cite{Patrick A. Lee} Different types of charge
inhomogeneity can emerge as a result of doping. Nanoscale inhomogeneity due to strong correlations between electrons, such as a stripe or
checkerboard order\cite{Emery,Zaanen89}, received a lot of attention in
recent years. \cite{Tranquada, Abbamonte, T. Hanaguri}  Other
sources of inhomogeneity are randomly
distributed heterovalent substituents or extra oxygen atoms, which introduce doped holes or electrons into the copper oxygen planes.\cite{Eisaki,Fratini} In
most cases (with some notable exceptions, such as Ortho
II-ordered YBa$_2$Cu$_3$O$_{6.5}$), the dopants act as charged
impurities. They form an inhomogeneous Coulomb potential
impacting the conduction electrons in the copper oxygen planes, resulting in an inhomogeneous charge distribution. Local
probes such as  scanning tunneling microscopy
(STM) have provided evidence for this inhomogeneity in cuprates such as
Bi$_2$Sr$_2$CaCu$_2$O$_{8+x}$ (BSCCO),\cite{S. H. Pan,McElroy,Fang,Kenjiro K. Gomes} Ca$_{2-x}$Na$_x$CuO$_2$Cl$_2$,\cite{Y. Kohsaka} and
La$_{2-x}$Sr$_x$CuO$_4$ (LSCO).\cite{T. Kato}  Bulk-sensitive nuclear magnetic resonance (NMR) and nuclear
quadropole resonance (NQR)
experiments confirmed that an inhomogeneous charge distribution is
not just a property of the surface, but is a feature of the
bulk in LSCO.\cite{P. M. Singer1,P. M. Singer2,Oleg}

Here we show how some doping-dependent features of the observed spectra of
phonons and electronic quasiparticles in LSCO, can be naturally
explained in terms of an inhomogeneous charge distribution. For
phonons, we base our analysis on new results of neutron scattering
experiments presented here. We use
previously published angle resolved photoemission (ARPES) data
for the electronic response.

It is a long-standing observation that the energy of the zone
boundary Cu-O bond-stretching (half-breathing) phonon, ${\bf q}=(0.5,0,0)$,
depends strongly on doping. It changes from about 81 meV at $x=0$ to
about 67 meV at $x=0.3$.\cite{L. Pintschovius,Pintschovius2,Dmitry2} Figure 1(b), which combines previous neutron scattering and x-ray scattering
data with our new results for the  frequency of the half-breathing mode, summarizes the available data.  
Since the phonon frequency depends strongly
on the doping, especially in the underdoped region,
an inhomogeneous doping should increase
its observed line width. We have measured the line width of this phonon in the
underdoped region to investigate this effect.

\begin{figure}
\centering \epsfxsize=8.0cm \epsfbox{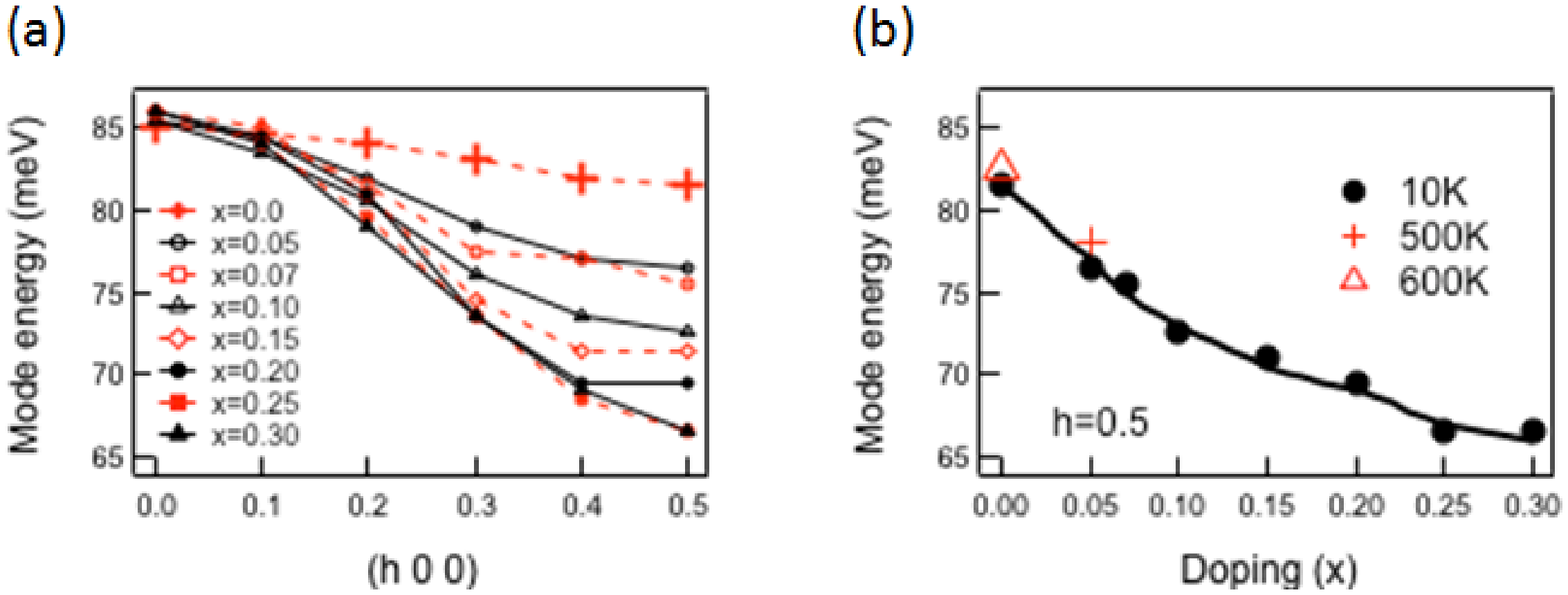} \caption{Doping
dependence of the Cu-O bond-stretching mode energy. a) Dispersion in the [100]-direction at different doping levels. b) Zone boundary half-breathing ($h=0.5$) mode energy as a function of doping (x).}\label{fig1}
\end{figure}

\section{Experimental Details}
Inelastic neutron scattering experiments were performed on  large, high-quality single crystals of LSCO with $x=0$, 0.05, 0.07,
and 0.15 on the 1T triple axis spectrometer at the ORPHEE reactor at
the Laboratoire Leon Brillouin at Saclay, France. The
monochromator and analyzer were the (220) reflection of Copper and
the (002) reflection of pyrolytic graphite (PG), respectively. The measurements
were performed at reciprocal lattice vectors ${\bf Q}=(4+h, 0,0)$ in
the tetragonal notation. In order to measure the intrinsic phonon
line width, it is important to properly account for the
experimental resolution, which depends not only on the instrument
configuration, but also on the phonon dispersion in the vicinity
of the measured point in ${\bf Q}$-$\omega$ space. We calculated the effective
energy-resolution for the bond-stretching phonon branch at each
doping level based on the spectrometer resolution combined with
doping-dependent dispersions calculated from the shell model,
whose parameters were optimized to agree with previously
published phonon dispersion data along high symmetry directions.\cite{Chaplot} The calculated experimental resolutions at Q=(4.5, 0, 0) for x=0.0, x=0.05, x=0.07 and x=0.15 are 4.5, 3.6, 3.2 and 3.4 meV, respectively. Electron-phonon coupling or anharmonicity should result in a Lorentzian intrinsic phonon line shape, but we found that a Gaussian gives better fits. Thus, the resolution function was convolved with a Gaussian function representing the phonon peak,
with the Gaussian width (and frequency and amplitude) determined by fitting to a constant-${\bf Q}$ scan. The background was determined from the scans at nearby wave vectors as was done in Ref. \onlinecite{Dmitry3}

\begin{figure}
\centering \epsfxsize=8.0cm \epsfbox{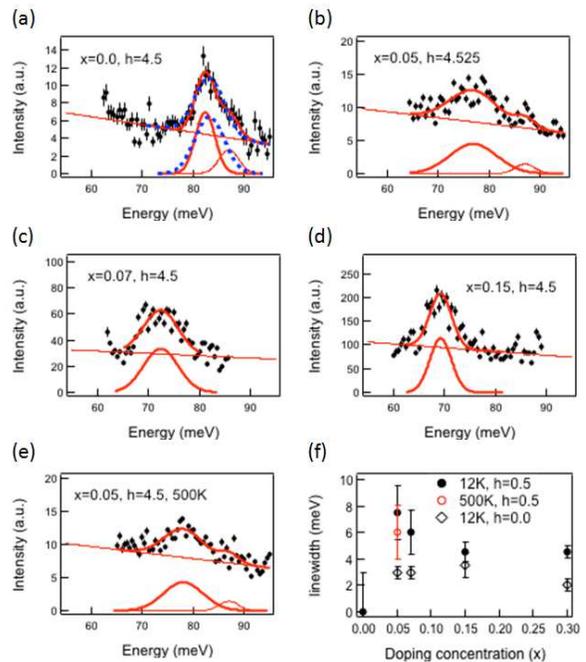} \caption{ 
Inelastic neutron scattering data of half breathing zone boundary
bond-stretching modes in LSCO. a) $x=0$, b) $x=0.05$, c) $x=0.07$, d) $x=0.15$
at 10~K, and e) $x=0.05$ at 500~K. Red lines on raw data are composed of a linear background and two Gaussian peaks for (a) and (b) and a linear
background and a Gaussian peak for (c), (d) and (e). Large Gaussian peaks on the bottom represent longitudinal bond-stretching modes; small Gaussian peaks at 87 meV represent the a suspect artifact (see text). Full widths of half maximum (FWHM) of the main peaks of x=0, x=0.05, x=0.07 and x=0.15 are 4.5 meV, 10.5 meV, 8 meV and 5meV, respectively. Two Gaussian peaks of x=0.0 data and a small Gaussian
peak of $x=0.05$ data are resolution-limited (4.5 meV). Dotted blue line on $x=0.0$ raw data is composed of a linear background and a Gaussian peak with 6.5 meV FWHM. f) Doping
dependence of the zone boundary ($h=0.5$) and zone center ($h=0.0$) longitudinal
bond-stretching mode line widths after deconvolving the resolution
function.}\label{fig2}
\end{figure}

\section{Results and discussion}

Figure 2(a-d) shows inelastic neutron scattering
(INS) spectra of the half-breathing mode
of LSCO from $x=0$ to $x=0.15$. The frequency clearly softens with increasing doping,
as previously reported.\cite{L. Pintschovius}  Red lines through
the raw data indicate the fitted Gaussian peak, convolved with the resolution function, on top of a linear background. In the cases of $x=0.0$ and $x=0.05$, we have included a second Gaussian to account for a small peak at 87 meV that we believe corresponds to an artifact,\cite{artifact} based on its lack of dependence on doping, although we do not have a definitive explanation for it.  We will ignore it in the following discussion of the line width; including it in the evaluation of the line width would not make a qualitative difference to the argument, as discussed below.

The fitted phonon line widths are plotted as a function of doping in Fig.~2(f).  The filled symbols show the width of the half-breathing mode at low temperature.  As one can see,  the width makes a large jump from $x=0$ to $x=0.05$, and then gradually decreases with further doping.  For comparison, the open diamonds show the fitted width of the zone-center mode.  The absence of any significant variation with doping suggests that the results for $h=0.5$ are intrinsic and are not a result of difference in sample quality. Furthermore, different samples with the same doping give identical results.

The most obvious explanation of phonon
broadening at the zone boundary is electron-phonon coupling. According to the conventional
theory of metals, the phonon line width due to electron-phonon
coupling  should be proportional to the density of states near
$E_F$,\cite{Allen, Grimvall} which is roughly proportional to doping in
LSCO.\cite{Yoichi Ando2, S. Uchida, Yoshida} The trend for the $h=0.5$ mode in Fig.~2(f) clearly violates a simple monotonic increase with doping.  In particular, the
$x=0.05$ sample is insulating\cite{Yoichi Ando1} and should not
have more conduction electrons near the Fermi surface than the
$x=0.07$ and 0.15 samples. Thus, we can rule out electron-phonon coupling as the dominant cause of this effect.

Another possible explanation of the broadening of the
low-temperature line shape is the effect of the tilt of the
CuO$_6$ octahedra responsible for the transition from the high-temperature tetragonal to the low-temperature orthorhombic phase.\cite{Yamada}  We
can rule out this scenario because octahedral tilt is largest for $x=0$ and decreases with doping.

\begin{figure}
\centering \epsfxsize=8.0cm \epsfbox{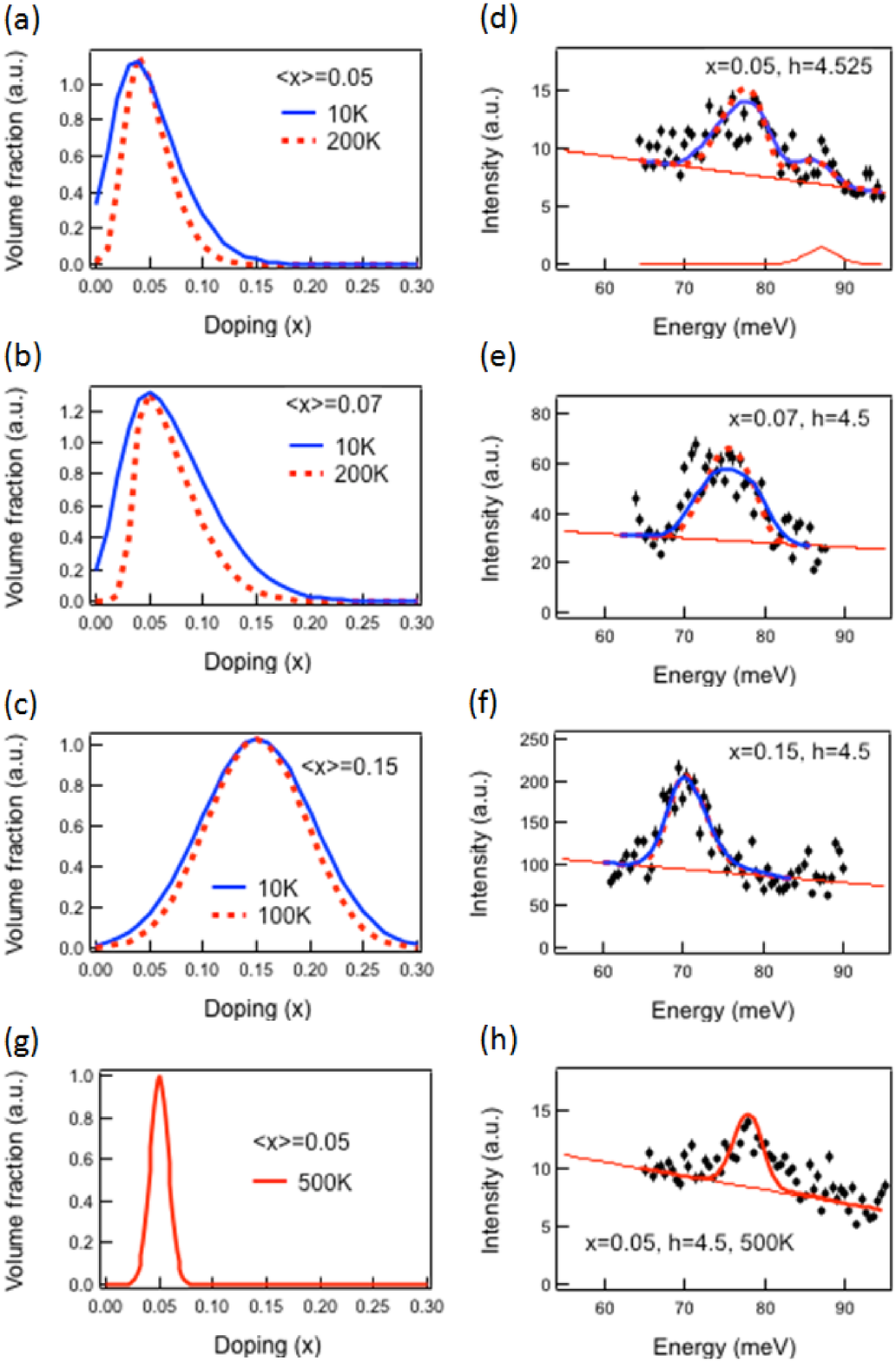} \caption{Doping distributions for $x=0.05$ (a and g), $x=0.074$ (b) and $x=0.15$ (c) based on the NQR data.\cite{P. M. Singer1} Inelastic neutron scattering data from the half-breathing mode for $x=0.05$ (d), $x=0.07$ (e) and $x=0.15$ (f) at 10 K and $x=0.05$ at 500~K (h). Dotted red curves and solid blue curves over data points show results calculated from the inhomogeneous doping effect at high temperature and at 10~K, respectively. Red curve in (h) shows the simulated results at 500~K.}\label{fig3}
\end{figure}

Here we propose a different mechanism to explain
our results. As already mentioned, NQR and NMR experiments found that the doping
level in nominally homogeneous samples of the LSCO family is
strongly spatially inhomogeneous.\cite{P. M. Singer1,P. M.
Singer2} Since the phonon frequency is known to depend on doping,
as shown in Fig.~1, we propose that the observed broadening
results from the distribution of local environments observed by
NQR. To test this idea we modeled phonon line shapes based on the
distribution of dopings measured by NQR\cite{P. M. Singer1}
combined with the doping dependence of phonon frequencies. 
(As the intrinsic phonon line-width is smaller than the resolution, we have ignored it in the modeling.) The
resulting line shapes were convolved with the resolution
function. This analysis provides a very rigorous test of the
above hypothesis, since it contains no adjustable parameters
aside from the overall peak magnitude.

One can imagine more rigorous approaches to calculate the phonon spectrum in an inhomogeneous medium such as calculations based on the coherent potential approximation (CPA) which is usually used for calculating phonon spectra in alloys.\cite{Gregg} But it is not clear if this method would apply to phonons in solids with charge inhomogeneity as in LSCO, since the CPA was originally developed for random, mass disordered alloys.\cite{Gregg} However, we note that calculations of disorder effects in isotopically 
disordered Ge using CPA gave no better results (actually even slightly 
inferior results) than calculations using large supercells.\cite{Gobel} 
Somewhat surprisingly, the supercell approach reproduced the 
experimentally observed phonon frequencies, linewidths and intensities 
very well, in spite of the fact that this approach is based on harmonic 
lattice dynamics. Therefore we think that our ansatz for calculating 
linewidths in LSCO - which is equivalent to a simplified version of the 
supercell approach - is quite adequate.


Figure 3 shows that the calculated line shapes reproduce the experimental low temperature phonon
line shapes at three doping levels reasonably well.  We also
measured this phonon in the $x=0.05$ sample at 500~K. Heating to such
temperatures increases anharmonicity and thus broadens the phonon. For example, this broadening has been observed
already at 330~K near the zone center in similar samples.\cite{D. Reznik} However,
the 500-K data appear to be slightly narrower than at 10~K, which
is consistent with decreasing charge inhomogeneity with increasing
temperature as reported by NQR.\cite{P. M. Singer1} 
An anharmonic contribution must be present, but it appears to be comparable to the contribution from charge inhomogeneity.

It is not completely clear why charge inhomogeneity
decreases upon heating the sample. One possibility is that some
electrons are trapped by the disordered chemical potential due to
randomly distributed Sr cations.\cite{P. M. Singer1} Then, the real space
distribution of the trapped electrons should broaden with
increasing temperature resulting in a smoother disorder
potential.\cite{P. M. Singer1} Charge inhomogeneity resulting from
electronic correlations should also decrease at elevated
temperatures due to the breakup of electronic self-organization
by thermal fluctuations.

Based on the above evidence we conclude that the phonon linewidth
can be well explained by charge inhomogeneities measured by NQR.
Intrinsic phonon lifetime, which is governed by anharmonicity and
electron-phonon coupling, is too long to measurably
contribute to the linewidth in addition to the effect of
inhomogeneous doping. 

\begin{figure}
\centering \epsfxsize=8.0cm \epsfbox{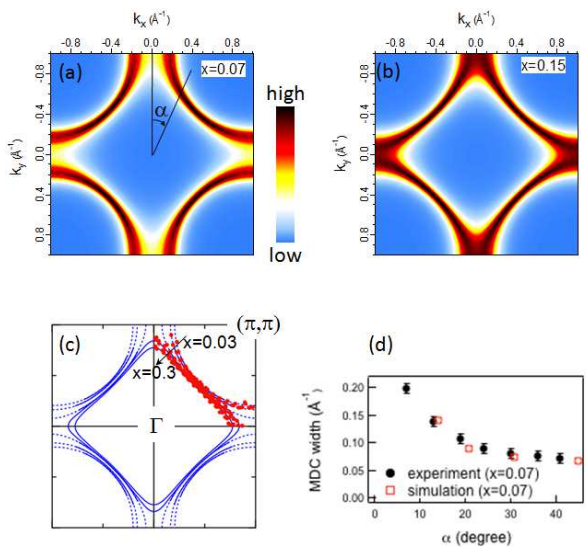} \caption{Simulated
Fermi surface results for (a) $x=0.07$ and (b) $x=0.15$.  (c) Doping
dependent Fermi surface evolution extracted from the ARPES data.\cite{Yoshida}
 (d) MDC widths at Fermi energy from experiment\cite{Yoshida2} and
simulation from anti-nodes to nodes.}\label{fig4}
\end{figure}

If this explanation is correct, charge
inhomogeneity should show up in other experiments involving spatially-averaged probes.  
In particular, electronic quasiparticles measured by ARPES should
also be affected by inhomogeneous doping, since the Fermi surface
in LSCO has a strong and nontrivial doping dependence: The Fermi
wave vector ${\bf k}_f$ is nearly doping-independent at the nodes, but
rapidly decreases with increased doping near the antinodes due to
the proximity to a Van Hove singularity (Fig.~4c).\cite{Yoshida,
Ino}  Similarly, measurements of the momentum width of the spectral function obtained from momentum distribution curves (MDCs) show that the width at the Fermi level is substantially larger in the antinodal region compared to the nodal one.\cite{Yoshida2,Valla00} We find that combining the chemical-potential-distribution model with the observed doping dependence of ${\bf k}_f$ provides a reasonable explanation for the ${\bf k}$ dependence of the momentum widths.

Figure 4(c) shows the doping-dependent Fermi surfaces
extracted from the ARPES data\cite{Yoshida}. We simulate the ARPES data by introducing a Lorentzian
broadening in energy of $\Delta E=$ 70 meV FWHM, which translates into $\Delta E/v_F$, broadening in {\bf k},  at each {\bf k}-point and assuming
a distribution of the electronic band structures weighted by the inhomogeneous
doping distribution, as we had done for the zone boundary phonon. 70 meV FWHM is a reasonable value based on the measured quasiparticle peak width near the Fermi energy in LSCO.\cite{Yoshida3} We use tight binding bands with first, second and third neighbor hopping parameters fitted to the ARPES data as the electronic band structures for each doping.\cite{Yoshida2} 
Figure 4(a) and (b) show the simulated results for the Fermi surfaces of
$x=0.07$ and $x=0.15$, respectively. Momentum widths of the quasiparticles near the Fermi surface broaden on going
from the nodal to the anti-nodal directions for both dopings, as
observed in experiments.\cite{X. J. Zhou} Figure 4(d)
demonstrates a good quantitative agreement between the MDC widths
calculated from our model and the data\cite{Yoshida2} for $x=0.07$. The line width of ARPES data at the Fermi energy has been, so far, usually interpreted as the self-energy effect of the quasiparticle due to impurity scattering. There is, however, no reason that the line width near anti-nodal direction is much broader than the line width near nodal direction due to impurity scattering which is believed to be isotropic. Our new analysis, yet, provides a natural explanation about broader line width near anti-nodal direction.

An alternative model is to apply a broadening of fixed width in momentum rather than energy; however, it was found that uniform momentum
broadening gives a much poorer agreement with the data. The physical significance of this result is unclear.

Broader ARPES line widths near the anti-nodal directions have been
interpreted in terms of stronger scattering,\cite{X. J. Zhou} but
we quantitatively reproduced the experimental data with a large
enhancement of the MDC widths towards the anti-nodal direction by
the inhomogeneous doping effect. Our simulation, which only includes doping inhomogeneity, cannot reproduce the entire ARPES spectrum, especially in the anti-nodal region, because of the pseudogap and the incoherent spectrum at higher binding energies.\cite{Meevasana}  However, we can conclude that the long-standing observation of broader peaks near anti-nodes in LSCO is due, in large part, to inhomogeneous doping.
We find further evidence
for the mechanism proposed here from another cuprate superconductor.
Quantum oscillation measurements showed that an overdoped Tl$_2$Ba$_2$CuO$_{6+d}$ (Tl2201) has  a much more
homogenous chemical potential,\cite{A. F. Bangura} because
the dopants in Tl2201 are farther from the copper oxide plane than
the dopants in LSCO and BSCCO.\cite{Eisaki} Recent ARPES data on this material show that the spectral functions
near the anti-nodes are even sharper than those at the nodes.\cite{M. Plate} These results are consistent with the conclusion that the broader lineshape near the anti-node than near the node is not universal in the cuprates but is a material dependent effect resulting from doping inhomogeneities.

Crystal momentum is generally a good quantum number for the electronic quasiparticles, whereas the length
scale of charge inhomogeneities in real space is only a few
nanometers.\cite{P. M. Singer1, S. H. Pan,Y. Kohsaka,T.
Kato,Kenjiro K. Gomes} Thus, it may be surprising that the spectra of phonons and of the electronic
quasiparticles are well reproduced by a linear sum of spectral
peaks for each doping weighted by the inhomogeneous doping
distribution function. For the electronic quasiparticles, 70 meV
FWHM intrinsic line width, combined with the measured Fermi velocity, translates into about 3 nm mean free
path for the nodal direction, which is almost the same as the length
scale of doping variation from NQR\cite{P. M. Singer1} and STM.\cite{T. Kato} It is even shorter for the
antinodal quasiparticles, because they have a smaller Fermi
velocity. We can, therefore, think that  electronic quasiparticles reside in each patch,
and simply add all spectral peaks from corresponding patches with different dopings.

For the phonon, we do not need additional line width broadening to fit the
experimental data as
shown in Fig.~3. This, however, does not mean that the phonon
has a much longer correlation length than the electronic
quasiparticles. If the wavelength of any waves in an inhomogeneous medium
is smaller than the length scale of the inhomogeneity of the medium, the
waves tend to be localized inside a domain of
the inhomogeneous medium.\cite{Michael,Nagel, Sajeev John, Tal Schwartz} The wavelength of zone
boundary phonons is two unit cells, which is about 4 times shorter than the
length scale of the inhomogeneity measured by STM and NQR.\cite{T. Kato, P. M. Singer1} Thus the
distribution of the zone boundary phonon frequencies in a medium
with inhomogeneous force constants directly reflects the distribution of the force constants.

To summarize, we have found evidence that elementary excitations in
LSCO such as phonons and electronic quasi-particles, are strongly
influenced by inhomogeneous doping previously reported by purely
local probes, NQR, NMR, and STM. Phonon and electronic spectra are well reproduced by just averaging over spectra
from patches having various doping concentrations, even though
the typical length scale of a patch is a few nanometers.
Good agreement between the model based on inhomogeneous doping
with both ARPES and neutron results
despite almost no adjustable parameters, provides compelling
evidence that inhomogeneous doping must be included into any
analysis of experimental results in the cuprates. Recently, iron pnictide supercondutors have also turned out to have chemical potential inhomogeneities due to a random dopant distribution.\cite{Yeoh} It is, therefore, necessary to take account of these charge inhomogeneities when interpreting measurements on iron pnictide superconductors as well. 

Inhomogeneity most likely originates from a random potential created by
introduced dopants (in LSCO it would be Sr), rather than from the
self-organization of conduction electrons, for example into
stripes. It is, yet, essential to gain further insight
into interactions of dynamic stripes with phonons, which is at
the focus of a different project. If stripes interact with
phonons via a mechanism discussed in Ref.~\onlinecite{D. Reznik, Eiji Kaneshita}, then the effect of dynamic stripes should occur at a
different wave vector, ${\bf q}=(0.25,0,0)$, not at the zone boundary. In
fact we find that the huge phonon line widths reported near this
wave vector in Ref. \onlinecite{D. Reznik} cannot be reproduced by inhomogeneous
doping alone for $x \geq 0.07$, and large additional
broadening is required near optimal doping. These results will be presented
in a subsequent publication.

\acknowledgements
The work at the University of Colorado was supported by the DOE, Office of Basic 
Energy Sciences under Contract No. DE-SC0006939. GDG and JMT are supported at Brookhaven by the Office of Basic Energy Sciences, Division of Materials Science and Engineering, U.S. Department of Energy (DOE), under Contract No. DE-AC02-98CH10886. The work at Tohoku University was supported by the Grant-In-Aid for Science Research A (22244039) from the MEXT of Japan.


\begin{thebibliography}{24}


\bibitem{Patrick A. Lee} Patrick A. Lee, Naoto Nagaosa and Xiao-Gang Wen, Rev. Mod. Phys. {\bf 78}, 17 (2006).
\bibitem{Emery} V. J. Emery, S. A. Kivelson, and H. Q. Lin, Phys. Rev. Lett. {\bf 64}, 475 (1990).
\bibitem{Zaanen89} J. Zaanen and O. Gunnarsson, Phys. Rev. B {\bf 40} 7391 (1989).
\bibitem{Tranquada} J. M. Tranquada, B. J. Sternlieb, J. D. Axe, Y. Nakamura and S. Uchida, Nature {\bf 375}, 561 (1995).
\bibitem{Abbamonte} P. Abbamonte, A. Rusydi, S. Smadici, G. D. Gu, G. A. Sawatzky and D. L. Feng, Nat. Phys. {\bf 1}, 155 (2005).
\bibitem{T. Hanaguri} T. Hanaguri, C. Lupien, Y. Kohsaka, D.-H. Lee, M. Azuma, M. Takano, H. Takagi and J. C. Davis, Nature {\bf 430}, 1001 (2004).
\bibitem{Eisaki} H. Eisaki, N. Kaneko, D. L. Feng, A. Damascelli, P. K. Mang, K. M. Shen, Z.-X. Shen and M. Greven, Phys. Rev. B {\bf 69}, 064512 (2004).
\bibitem{Fratini} Michela Fratini, Nicola Poccia,	 Alessandro Ricci, Gaetano Campi, Manfred Burghammer, Gabriel Aeppli and Antonio Bianconi, Nature {\bf 466}, 841 (2010).
\bibitem{S. H. Pan} S. H. Pan, J. P. O'Neal, R. L. Badzey, C. Chamon, H. Ding, J. R. Engelbrecht, Z. Wang, H. Eisaki, S. Uchida, A. K. Gupta, K.-W. Ng, E. W. Hudson, K. M. Lang and J. C. Davis, Nature {\bf 413}, 282 (2001).
\bibitem{McElroy} K. McElroy, J. Lee, J. A. Slezak, D.-H. Lee, H. Eisaki, S. Uchida, and J. C. Davis, Science {\bf 309}, 1048 (2005).
\bibitem{Fang} A. C. Fang, L. Capriotti, D. J. Scalapino, S. A. Kivelson, N. Kaneko, M. Greven, and A. Kapitulnik, Phys. Rev. Lett. {\bf 96}, 017007 (2006).
\bibitem{Kenjiro K. Gomes} Kenjiro K. Gomes, Abhay N. Pasupathy, Aakash Pushp, Shimpei Ono, Yoichi Ando and Ali Yazdani, Nature {\bf 447}, 569 (2007).
\bibitem{Y. Kohsaka} Y. Kohsaka, K. Iwaya, S. Satow, T. Hanaguri, M. Azuma, M. Takano and H. Takagi, Phys. Rev. Lett. {\bf 93}, 097004 (2004).
\bibitem{T. Kato} T. Kato, S. Okitsu and H. Sakata, Phys. Rev. B {\bf 72}, 144518 (2005).
\bibitem{P. M. Singer1} P. M. Singer, A. W. Hunt, and T. Imai, Phys. Rev. Lett. {\bf 88}, 047602 (2002).

\bibitem{P. M. Singer2} P. M. Singer, T. Imai, F. C. Chou, K. Hirota, M. Takaba, T. Kakeshita, H. Eisaki and S. Uchida, Phys. Rev. B {\bf 72}, 014537 (2005).
\bibitem{Oleg} Wei Chen, Giniyat Khaliullin and Oleg P. Sushkov, Phys. Rev. B {\bf 80}, 094519 (2009).
\bibitem{L. Pintschovius} For a review of early results see L. Pintschovius, Phys. Stat. Solids (b) {\bf 242}, No. 1, 30Ð 50 (2005).
\bibitem{Pintschovius2} L. Pintschovius, D. Reznik and K. Yamada, Phys. Rev. B {\bf 74}, 174514 (2006).
\bibitem{Dmitry2} D. Reznik, Advances in Condensed Matter Physics {\bf 2010}, Article ID 523549 (2010).
\bibitem{Chaplot} S. L. Chaplot, W. Reichardt, L. Pintschovius and N. Pyka, Phys. Rev. B {\bf 52}, 7230 (1995).
\bibitem{Dmitry3} D. Reznik, L. Pintschovius, M. Fujita, K. Yamada, G. D. Gu and J. M. Tranquada, Journal of Low Temperature Physics {\bf 147}, 353Ð364 (2007)
\bibitem{artifact}{We note that the total phonon density of states\cite{L. Pintschovius,R. J. McQueeney} has a peak at 87 meV that is relatively insensitive to doping; however, it would require an unreasonable amount of polycrystalline material in our samples for that to provide an explanation.  We cannot rule out a multi-phonon effect.}
\bibitem{R. J. McQueeney} R. J. McQueeney, J. L. Sarrao, P. G. Pagliuso, P. W. Stephens and R. Osborn, Phys. Rev. Lett. {\bf 87}, 077001 (2001).
\bibitem{Allen} P.B. Allen, Solid State Commun. {\bf 14}, 937 (1974)
\bibitem{Grimvall} G. Grimvall, The Electron-Phonon Interaction in Metals, Selected Topics in Solid State Physics, edited by E. Wohlfarth (North-Holland, New York, 1981).
\bibitem{Yoichi Ando2} S. Ono, Seiki Komiya, and Yoichi Ando, Phys. Rev. B {\bf 75}, 024515 (2007).
\bibitem{S. Uchida} S. Uchida, T. Ido, H. Takagi, T. Arima, Y. Tokura and S. Tajima, Phys. Rev. B {\bf 43}, 7942 (1991).
\bibitem{Yoshida} T. Yoshida, X. J. Zhou, K. Tanaka, W. L. Yang, Z. Hussain, Z.-X. Shen, A. Fujimori, S. Sahrakorpi, M. Lindroos, R. S. Markiewicz, A. Bansil, Seiki Komiya, Yoichi Ando, H. Eisaki, T. Kakeshita and S. Uchida, Phys. Rev. B {\bf 74}, 224510 (2006).
\bibitem{Yoichi Ando1} Yoichi Ando, A. N. Lavrov, Seiki Komiya, Kouji Segawa and X. F. Sun, Phys. Rev. Lett. {\bf 87}, 017001 (2001).
\bibitem{Yamada} K. Yamada, C. H. Lee, K. Kurahashi, J. Wada, S. Wakimoto, S. Ueki, H. Kimura, Y. Endoh, S. Hosoya, G. Shirane, R. J. Birgeneau, M. Greven, M. A. Kastner, and Y. J. Kim, Phys. Rev. B {\bf 57}, 6165 (1998).
\bibitem{Gregg} J. R. Gregg and C. W. Myles, J. Phys. Chem. Solids {\bf 46}, 1305 (1985).
\bibitem{Gobel} A. Go\"bel, D.T.Wang, M. Cardona, L. Pintschovius, W. Reichardt, J. Kulda, N.M.Pyka, K.Itoh and E.E. Haller, Phys. Rev. B {\bf 58}, 10510 (1998).
\bibitem{Rosch} O. Ro\"sch and O. Gunnarsson, Phys. Rev. Lett. {\bf 92}, 146403 (2004).

\bibitem{D. Reznik} D. Reznik, L. Pintschovius, M. Ito, S. Iikubo, M. Sato, H. Goka, M. Fujita, K. Yamada, G. D. Gu and J. M. Tranquada, Nature {\bf 440}, 1170 (2006).
\bibitem{Yoshida2} T. Yoshida, X J Zhou, D H Lu, Seiki Komiya, Yoichi Ando, H Eisaki, T Kakeshita, S Uchida, Z Hussain, Z-X Shen and A Fujimori, J. Phys.: Condens. Matter  {\bf 19}, 125209 (2007).
\bibitem{Ino} A. Ino, C. Kim, M. Nakamura, T. Yoshida, T. Mizokawa, A. Fujimori, Z.-X. Shen, T. Kakeshita, H. Eisaki and S. Uchida, Phys. Rev. B {\bf 65}, 094504 (2002).
\bibitem{Valla00} T. Valla, A. V. Fedorov, P. D. Johnson, Q. Li, G. D. Gu, and N. Koshizuka, Phys. Rev. Lett. {\bf 85}, 828 (2000).
\bibitem{Yoshida3} T. Yoshida, M. Hashimoto, S. Ideta, A. Fujimori, K. Tanaka, N. Mannella, Z. Hussain, Z.-X. Shen, M. Kubota, K. Ono, Seiki Komiya, Yoichi Ando, H. Eisaki and S. Uchida. Phys. Rev. Lett. {\bf 103}, 037004 (2009).



\bibitem{X. J. Zhou} X. J. Zhou, T. Yoshida, D.-H. Lee, W. L. Yang, V. Brouet, F. Zhou, W. X. Ti, J. W. Xiong, Z. X. Zhao, T. Sasagawa, T. Kakeshita, H. Eisaki, S. Uchida, A. Fujimori, Z. Hussain and Z.-X. Shen, Phys. Rev. Lett. {\bf 92}, 187001 (2004).
\bibitem{Meevasana} W. Meevasana, X. J. Zhou, S. Sahrakorpi, W. S. Lee, W. L. Yang, K. Tanaka, N. Mannella, T. Yoshida, D. H. Lu, Y. L. Chen, R. H. He, Hsin Lin, S. Komiya, Y. Ando, F. Zhou, W. X. Ti, J. W. Xiong, Z. X. Zhao, T. Sasagawa, T. Kakeshita, K. Fujita, S. Uchida, H. Eisaki, A. Fujimori, Z. Hussain, R. S. Markiewicz, A. Bansil, N. Nagaosa, J. Zaanen, T. P. Devereaux and Z.-X. Shen, Phys. Rev. B {\bf 75}, 174506 (2007).
\bibitem{A. F. Bangura} A. F. Bangura, P. M. C. Rourke, T. M. Benseman, M. Matusiak, J. R. Cooper, N. E. Hussey and A. Carrington, Phys. Rev. B {\bf 82}, 140501(R) (2010).
\bibitem{M. Plate} M. Plate, J. D. F. Mottershead, I. S. Elfimov, D. C. Peets, Ruixing Liang, D. A. Bonn, W. N. Hardy, S. Chiuzbaian, M. Falub, M. Shi, L. Patthey and A. Damascelli, Phys. Rev. Lett. {\bf 95}, 077001 (2005).
\bibitem{Michael} Michael J. O'Hara, Charles W. Myles, John D. Dow and Ronald D. Painter, J. Phys. Chem. Solids {\bf42}, 1043 (1981).
\bibitem{Nagel} Sidney R. Nagel, A. Rahman and Gary S. Grest, Phys. Rev. Lett. {\bf 47}, 1665 (1981).
\bibitem{Sajeev John} Sajeev John, H. Sompolinsky and Michael J. Stephen, Phys. Rev. B {\bf 27}, 5592 (1983).
\bibitem{Tal Schwartz} Tal Schwartz, Guy Bartal, Shmuel Fishman and Mordechai Segev, Nature {\bf 446}, 52 (2007).
\bibitem{Yeoh} W. K. Yeoh, B. Gault, X. Y. Cui, C. Zhu, M. P. Moody, L. Li, R. K. Zheng, W. X Li, X. L. Wang, S. X. Dou, G. L. Sun, C. T. Lin and S. P. Ringer, Phys. Rev. Lett. {\bf 106}, 247002 (2011).
\bibitem{Eiji Kaneshita} Eiji Kaneshita, Masanori Ichioka and Kazushige Machida, Phys. Rev. Lett. {\bf 88}, 115501 (2002).





\end{thebibliography}
\end{document}